%% file: conf_paper.tex
\documentclass[11pt]{article}
\usepackage{graphicx,rotating}
\usepackage{multirow}
\newcommand{\BABARPubYear}    {04}

\newcommand{\BABARConfNumber} {04}
\newcommand{\SLACPubNumber} {10593}

\input pubboard/babarsym
\input includes/particle.tex

\long\def\inst#1{\par\nobreak\kern 4pt\nobreak
    {\it #1}\par\vskip 10pt plus 3pt minus 3pt}

\begin{document}
{\pagestyle{empty}

\begin{flushright}

\babar-CONF-\BABARPubYear/\BABARConfNumber \\
SLAC-PUB-\SLACPubNumber \\

August 2004 \\
\end{flushright}
\par\vskip 5cm

\begin{center}
\Large \bf Measurements of $\Lambda^+_c$ Branching Fractions of Cabibbo-Suppressed Decay Modes.
\end{center}
\bigskip

\begin{center}
\large The \babar\ Collaboration\\
\mbox{ }\\
\today
\end{center}
\bigskip \bigskip

\begin{center}
\large \bf Abstract
\end{center}

We have measured the branching fractions of the Cabibbo-suppressed decays $\Lambda^+_c$ $\to$ \lz $K^+$ and $\Lambda^+_c$ $\to$ \sigz $K^+$
relative to the Cabibbo-allowed decay modes $\Lambda^+_c$ $\to$ \lz $\pi^+$ and $\Lambda^+_c$ $\to$ \sigz $\pi^+$ to be $ 0.044~ \pm ~0.004 ~(\textnormal{stat.})~ \pm ~0.002 ~(\textnormal{syst.})$ and $ 0.040~ \pm ~0.005 ~(\textnormal{stat.})~ \pm ~0.004 ~(\textnormal{syst.})$, respectively.
We also present the first observation of $\Lambda^+_c$ $\to$ \lz $K^+ \pi^+ \pi^-$ and have measured the branching fraction relative to $\Lambda^+_c$ $\to$ \lz $\pi^+$ to be $0.266~ \pm ~0.027 ~(\textnormal{stat.})~ \pm ~0.032 ~(\textnormal{syst.})$. The upper limit of the branching fraction into the decay $\Lambda^+_c$ $\to$ \sigz $K^+ \pi^+ \pi^-$ relative to  $\Lambda^+_c$ $\to$ \sigz $\pi^+$ has been measured to be 
$ < 3.9 \times ~10^{-2}$ at the 90\% confidence level. This analysis was performed using a data sample of 
 125 fb$^{-1}$ (integrated luminosity) collected by the \babar$\ $detector at the PEP-II asymmetric-energy $B$ Factory at the Stanford Linear Accelerator Center. 
All results presented in this conference contribution are preliminary.
\vfill
\begin{center}

Submitted to the 32$^{\rm nd}$ International Conference on High-Energy Physics, ICHEP 04,\\
16 August---22 August 2004, Beijing, China

\end{center}

\vspace{1.0cm}
\begin{center}
{\em Stanford Linear Accelerator Center, Stanford University, 
Stanford, CA 94309} \\ \vspace{0.1cm}\hrule\vspace{0.1cm}
Work supported in part by Department of Energy contract DE-AC03-76SF00515.
\end{center}

\newpage
} 

\input pubboard/authors_sum2004.tex

\section{INTRODUCTION}
\label{sec:Introduction}
~~~Beginning with the first observation of the charmed baryon $\Lambda_c^+$ in
1979 by MARK-II and BNL ~\cite{MARK,Samios}, our knowledge of the physics of charmed baryons developed less rapidly than that of the charmed mesons.
This is due to the smaller baryon production cross section, shorter life
time, and, in \epem storage rings, the absence of a cleanly observable
 $\Lambda_c$ $\bar{\Lambda}_c$ resonance. During the last few years there has been significant progress in the
experimental study of the hadronic decays of charmed baryons. Recent results
 on masses, widths, lifetimes, production rates and the decay asymmetry parameters have been
published by different experiments; among them 
the discoveries of Cabibbo-suppressed decays $\Lambda^+_c \to \p\phiz$ by CLEO~\cite{CLEO}, and $\Lambda^+_c \to \Lambda^{0}K^{+}$ , $\Lambda^+_c \to \Sigma^{0}K^{+}$ by Belle ~\cite{Belle}.

 The precision in the measurements of branching fractions is only about
 40$\%$ for many Cabibbo-favored modes~\cite{PDG}, while for 
Cabibbo-suppressed decays the precision is even
worse. As a consequence, we are not yet able to distinguish between
the decay rate predictions made by different models, e.g., the quark model approach to non-leptonic charm decays and the Heavy Quark Effective Theory(HQET)~\cite{Khana,Kram, Kron}.

  In this paper we present a study of $\Lambda_c^+$ baryons produced in the 
\epem $\to$ \qqbar continuum at \babar. We present improved measurements of the Cabibbo-suppressed decays $\Lambda^+_c$ $\to$  $\Lambda^{0} K^+$ and $\Lambda^+_c$ $\to$ $\Sigma^{0} K^+$, report the first observation of $\Lambda^+_c$ $\to$ $\Lambda^{0} K^+ \pi^+ \pi^-$, and set an upper limit on $\Lambda^+_c$ $\to$ $\Sigma^{0} K^+ \pi^+ \pi^-$. Here and throughout this 
paper, inclusion of charge-conjugate states is implied.

\section{THE \babar\ DETECTOR AND DATASET}
\label{sec:babar}

~~~The data sample used in this analysis consists of 125 \invfb integrated luminosity recorded between October 1999 and June 2003 with the \babar\ detector at the \slac\ \pep2\ storage ring. The \pep2\ facility operates nominally at the \FourS\ resonance, providing collisions of 9.0\gev electrons on 3.1\gev positrons. The data set includes 112 \invfb collected at the \FourS (``on-resonance'') and  13 \invfb collected below the $B \bar B$ threshold (``off-resonance'').   

A detailed description of the \babar\ detector can be found elsewhere~\cite{babar}; only detector components most relevant to this analysis are mentioned here. Charged-particle trajectories are measured by a five-layer double-sided silicon vertex tracker (SVT) and a 40-layer drift chamber (DCH), operating in the field of a 1.5-T solenoid. Charged particles are identified by combining measurements of ionization energy loss ($dE/dx$) in the DCH and SVT with angular information from a detector of internally reflected Cherenkov light (DIRC). Photons are identified as isolated electromagnetic showers in a CsI(Tl) electromagnetic calorimeter.

We have used Monte Carlo simulations (MC) of the \babar\ detector based on GEANT4~\cite{G4} to optimize our selection criteria and to determine signal efficiencies. These simulations take into account the varying detector conditions and beam backgrounds 
during the data-taking period.

\section{ANALYSIS METHOD}
\label{sec:Analysis}
For this analysis, the particle identification is important. 
A track is identified as a kaon, pion, or proton if it is projected to pass through the fiducial volume of the DIRC and the reconstructed cone of Cerenkov light is consistent in time and angle with the measured track momentum. This information is augmented with $dE/dx$ measured with the SVT and DCH.
Photons are detected in the CsI calorimeter.

\begin{figure}[!htb]
\begin{center}
\includegraphics[height=7cm]{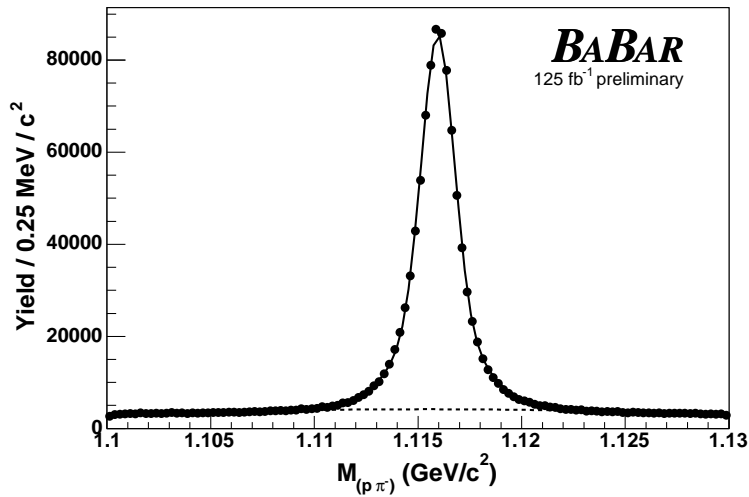}
\caption{The invariant mass of $p \pi$ combinations (GeV/$c^{2}$).}
\label{fig:lambda}
\end{center}
\end{figure}

\begin{figure}[!htb]
\begin{center}
\includegraphics[height=7cm]{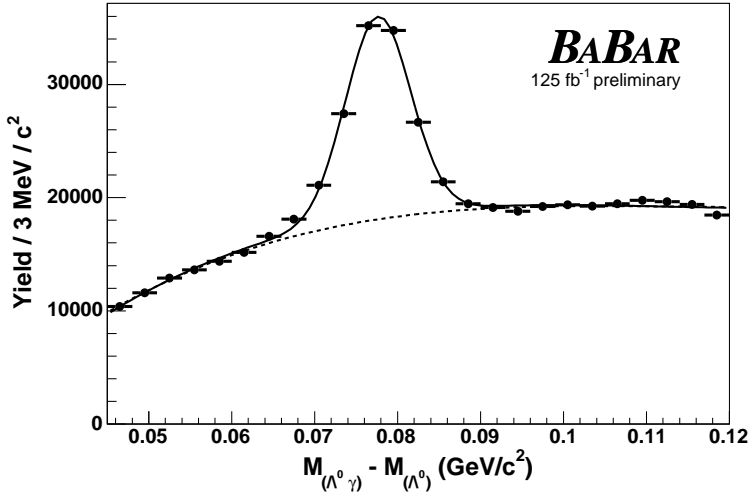}
\caption{The invariant mass difference between $\Lambda^{0} \gamma$ combinations and $\Lambda^{0}$ candidates (GeV/$c^{2}$).}
\label{fig:sigma}
\end{center}
\end{figure}

\indent Candidates for $\Lambda^{0}$ are reconstructed in the decay mode $\Lambda^{0} \to p \pi^{-}$.
 We fitted the $p$ and $\pi^{-}$ tracks to a common vertex and required the probability of 
$\chi^{2}$ of vertex fit to be greater than 0.1 $\%$. We also required the (three-dimensional) flight distance of each $\Lambda^{0}$ candidate between 
its decay vertex and the 
primary vertex
to be greater than 0.2~cm.
We fitted the invariant mass of $\Lambda^{0}$ candidates, with a double-Gaussian function of common mean to represent the signal and a second-order polynomial to represent the background. This fit is shown in Fig.~\ref{fig:lambda}. The fitted resolution is $\sigma= \sigma_{\rm {RMS}} = 1.5~\mevcsq$,
where $\sigma_{\rm {RMS}}$ is defined by
\[ \sigma_{\rm {RMS}}^{2} \equiv f_{1} \sigma_{1}^{2} + f_{2} \sigma_{2}^{2},\]
where $f_{1}$ and $f_{2}$ are fractions of signal yield corresponding to
Gaussian functions one and two, respectively, and $\sigma_{1}$ and $\sigma_{2}$ are the two corresponding widths. 
We required the mass of  $\Lambda^{0}$ candidates to be in the range 1113~\mevcsq~$<$ $M_{\Lambda^{0}}$ $<$ 1119~\mevcsq. 
The $\Sigma^{0}$ candidates were reconstructed in the decay mode $\Sigma^{0} \to  \Lambda^{0} \gamma $ using the $\Lambda^{0}$ sample and photons
 with a calorimeter cluster energy greater than 0.1~GeV. 
The mass difference($M_{\Lambda^{0} \gamma} - M_{\Lambda^{0}}$) is shown in Fig.~\ref{fig:sigma}. Fitting with two Gaussian functions of common mean for the signal contribution, and with a third-order polynomial for background, we obtained a resolution $\sigma$ = $\sigma_{\rm RMS} = 4.0~\mevcsq$ and a mean ( difference between the invariant masses) of 767 $\pm$ 1~\mevcsq. We accept
candidates with $M_{\Lambda^{0} \gamma} - M_{\Lambda^{0}}$ within $\pm$ 10.0~\mevcsq. (2.5 $\sigma$) of the measured mean value.
\newline
\indent To suppress combinatorial and $B\bar{B}$ backgrounds, we required
 $\Lambda^+_c$ candidates to have scaled momentum $x_{p}$ = $p^{*}$/$p^{*}_{\rm max} > $ 0.5; here $p^{*}$ is the reconstructed momentum of the $\Lambda^+_c$ 
candidate in the \epem center of mass, $p^{*}_{\rm max}$ = $\sqrt{s/4 - M^{2}}$, $\sqrt{s}$ is the total center of mass energy and $M$ is the reconstructed mass of the $\Lambda^+_c$ candidate. The signal detection efficiency is obtained from MC with particle identification efficiency corrections based on data.

\section{PHYSICS RESULTS}
\label{sec:Physics}

\subsection{Measurement of the decays $\Lambda^+_c \to \Lambda^{0}K^+$ and $\Lambda^+_c \to \Sigma^{0}K^+$}
\label{sec:Measurement of LS}
~~~The Cabibbo-suppressed decay $\Lambda^+_c \to \Lambda^{0}K^{+}$ 
was first measured 
by the Belle Collaboration~\cite{Belle}. 

\begin{figure}[!htb]
\begin{center}
\includegraphics[height=7cm]{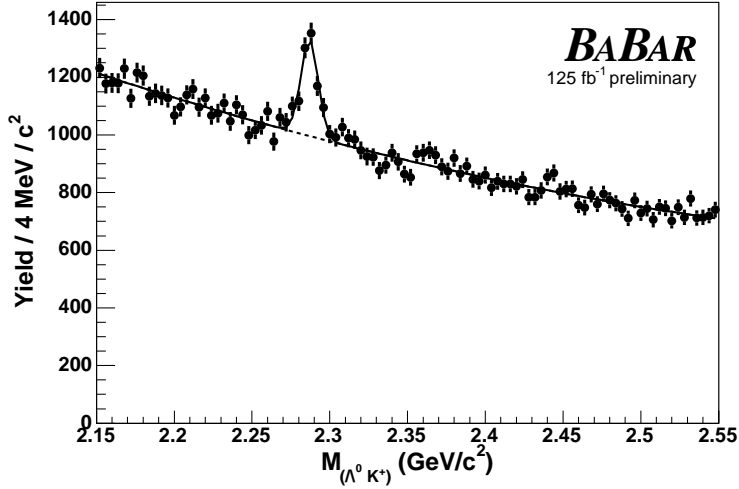}
\caption{The invariant mass of $\Lambda^{0}K^+$ combinations (GeV/$c^{2}$).}
\label{fig:lctolk}
\end{center}
\end{figure}

For our analysis, the $\Lambda^{0}$ and $K^+$ were combined to form $\Lambda^+_c$ as 
described in Sec.~\ref{sec:Analysis}. Invariant mass distribution of 
$\Lambda^{0}K^+$ combinations is shown in Fig.~\ref{fig:lctolk}. The mass 
distribution was fitted with a Gaussian function for the signal, and a second-order polynomial
 for combinatorial background. We obtain a raw yield of 1164 $\pm$ 107 (\textnormal{stat.}) events, with a 10.9 standard deviations significance and with fitted width $\sigma$  = 5.5 $\pm$ 0.7~\mevcsq, consistent with the MC prediction of 6.0~\mevcsq.

\begin{figure}[!htb]
\begin{center}
\begin{tabular}{c}

\mbox{\includegraphics[height=9.5cm,width=7.5cm,angle=90]{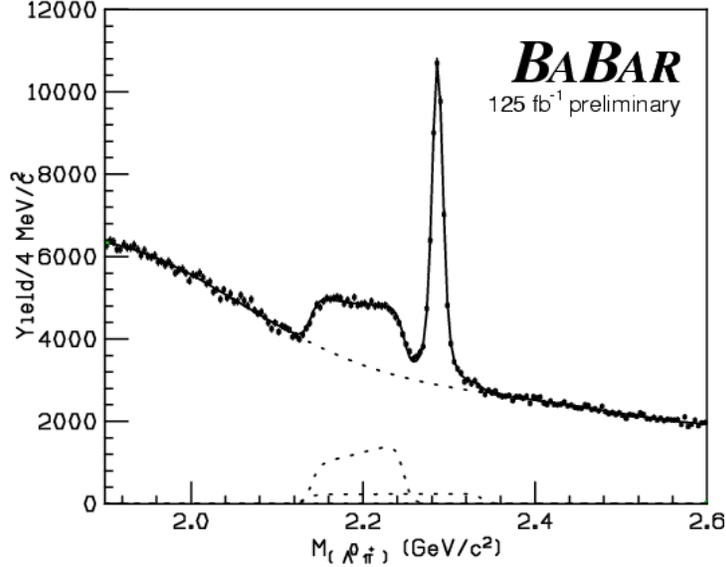}}
\end{tabular}
\caption{The invariant mass of $\Lambda^{0}\pi^+$ combinations (GeV/$c^{2}$).}
\label{fig:lctolp}
\end{center}
\end{figure}
\begin{figure}[!htb]
\begin{center}
\includegraphics[height=7cm]{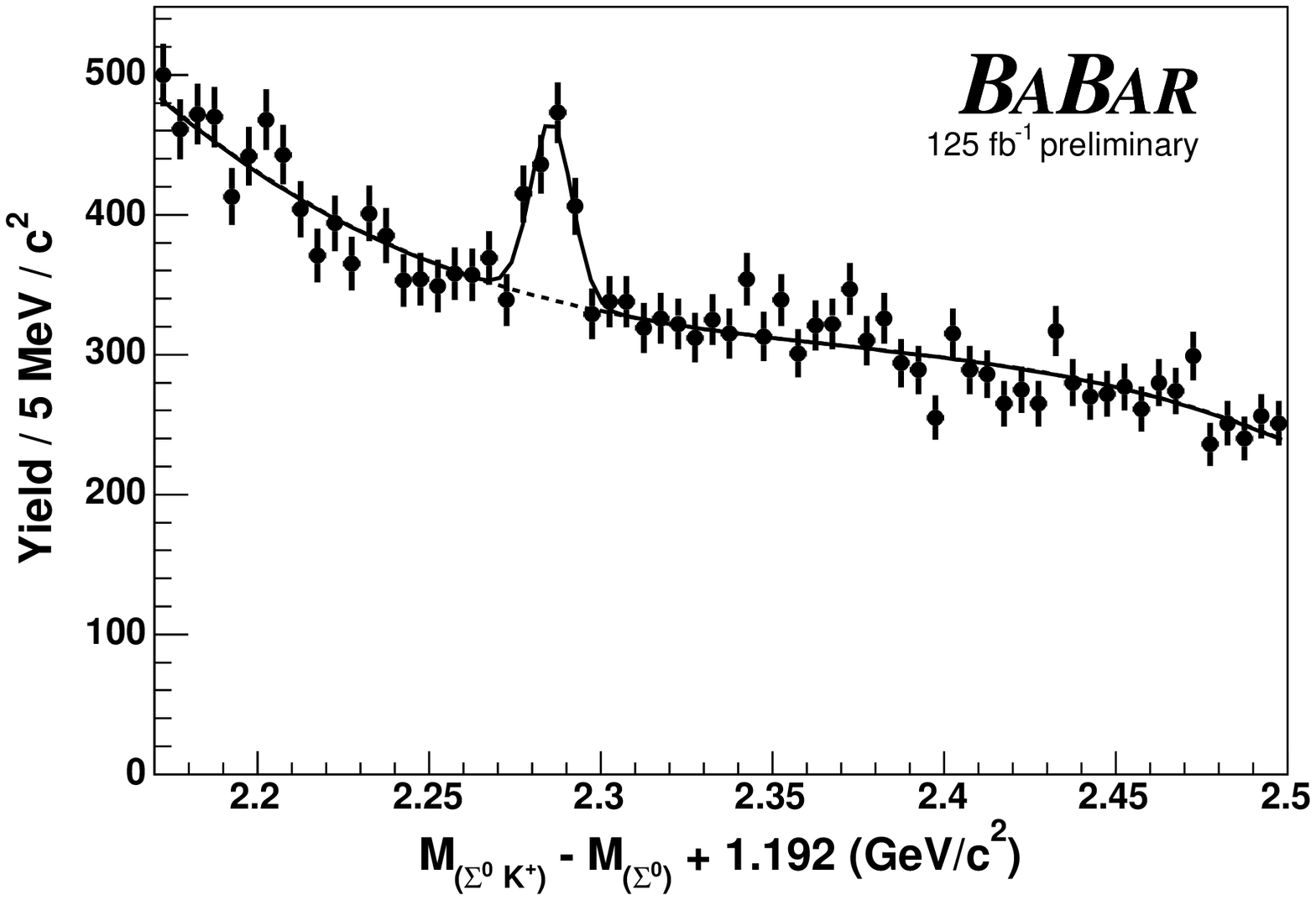}
\caption{The invariant mass of $\Sigma^{0}K^+$ combinations (GeV/$c^{2}$).}
\label{fig:lctosk}
\end{center}
\end{figure}
\begin{figure}[!htb]
\begin{center}
\begin{tabular}{c}
\mbox{\includegraphics[height=7cm]{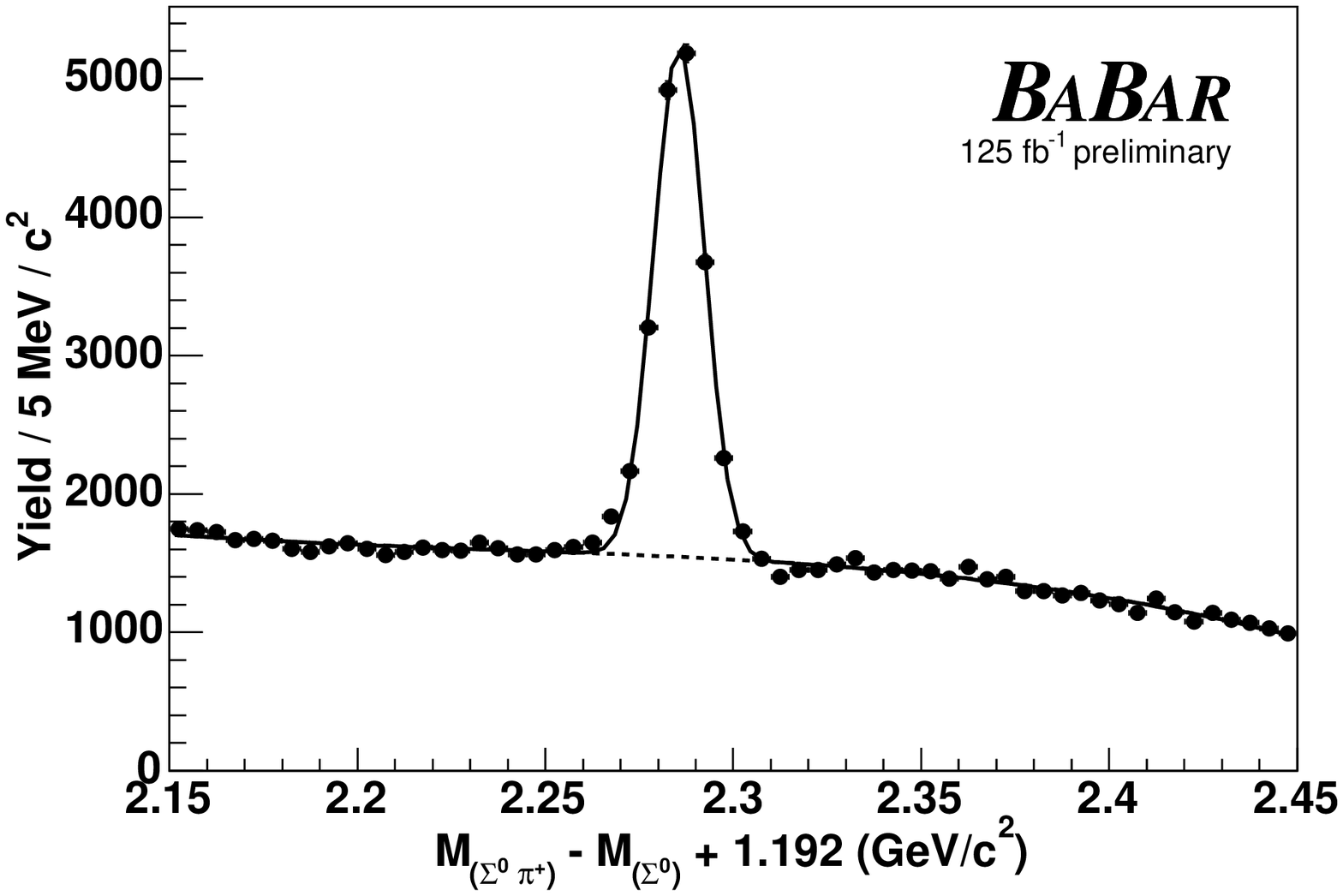}}
\end{tabular}
\caption{The invariant mass of $\Sigma^{0}\pi^+$ combinations (GeV/$c^{2}$).}
\label{fig:lctosp}
\end{center}
\end{figure}

\indent For normalization, we used the decay $\Lambda^+_c \to \Lambda^{0}\pi^{+}$. The invariant mass distribution of $\Lambda^{0}\pi^{+}$ combinations is
shown in Fig.~\ref{fig:lctolp}.
At mass values below the peak centered at the $\Lambda^+_c$ mass a broad distribution at 2.2 GeV/$c^{2}$ is visible which is a reflection due to 
$\Lambda^+_c \to \Sigma^{0}\pi^+$ with a missing $\gamma$. Additionally, at 2.3 GeV/$c^{2}$ we see a shoulder, the upper edge of a $\Xi_c$ reflection whose full shape extends through the entire $\Lambda^+_c$ signal region.
The distribution was fitted using two Gaussian functions with same mean for 
signal, a square wave function for each reflection, and a 7th-order polynomial
for combinatorial background. We obtained the width $\sigma = 8.2~\mevcsq$, which is consistent with the experimental resolution of about $8.0~\mevcsq$, and a raw 
yield of 33594 $\pm$ 367 (stat.) events. We calculate the ratio {\brc(\lcpl \ra \lz \kpl)}/{\brc(\lcpl \ra \lz \pipl )} in on-resonance and off-resonance data. The results are comparable within uncertainties, so 
off-resonance data is combined with on-resonance data. Using 
signal MC, the relative 
 signal reconstruction efficiency is found to be $\epsilon ( \Lambda^+_c \to \Lambda^{0}K^{+})$ / $\epsilon ( \Lambda^+_c \to \Lambda^{0}\pi^{+})$ = 0.781~ $\pm$ ~0.004 (stat.). With this value we calculate:
\[\frac{\brc(\lcpl \ra \lz \kpl)}{\brc(\lcpl \ra \lz \pipl )} = ~0.044~ \pm ~0.004 ~(\textnormal{stat.})~ \pm ~0.002 ~(\textnormal{syst.})~.\]
We provide a detailed description of the sources of systematic uncertainty for this
and other measured decay modes in Sec.~\ref{sec:Systematics}. \\
\indent  The Cabibbo-suppressed decay $\Lambda^+_c \to \Sigma^{0}K^{+}$ was first measured by the Belle collaboration~\cite{Belle}, under the restriction of the scaled momentum to the region $x_{p} > $ 0.6. For our analysis, we combined a $\Sigma^{0}$ and a $K^{+}$ candidate to form $\Lambda^+_c$, requiring $x_{p} > $ 0.5 as before. The 
invariant mass of $\Sigma^{0}K^{+}$ combinations is shown in 
Fig.~\ref{fig:lctosk}. We improved the invariant mass resolution by plotting the  corrected mass e.g., $M_{\Sigma^{0}K^{+}} - M_{\Sigma^0} + M^{PDG}_{\Sigma^0}$, instead of $M_{\Sigma^{0}K^{+}}$ where $M_{\Sigma^0}$ is the reconstructed mass of ${\Sigma^0}$ and $M^{PDG}_{\Sigma^0}$ is the mass of ${\Sigma^0}$ from PDG~\cite{PDG}.
We fit the distribution using a Gaussian function with a fixed width $\sigma$ = 6.0~\mevcsq\ (as determined from MC simulations) for signal and a third-order polynomial for combinatorial backgrounds. The fit yields 387 $\pm$ 48 (stat.) events with a 8.1$\sigma$ statistical significance for $\Lambda^+_c$ baryons having decayed to $\Sigma^{0}K^{+}$. For normalization, we 
use the Cabibbo-allowed decay mode $\Lambda^+_c \to \Sigma^{0}\pi^{+}$. The invariant mass
of $\Sigma^{0}\pi^{+}$ combinations is shown in Fig.~\ref{fig:lctosp}. The
fit uses a Gaussian function for the signal and a third-order polynomial for background.
The measured width $\sigma$ = 6.7 $\pm$ 0.1~\mevcsq, which is consistent with the MC prediction of $\sigma=7.0$~\mevcsq. The fitted yield is $12450~\pm~170$ (stat.) events. The relative reconstruction efficiency
is measured to be $\epsilon ( \Lambda^+_c \to \Sigma^{0}K^{+})$ / $\epsilon ( \Lambda^+_c \to \Sigma^{0}\pi^{+})$ = 0.780 ~$\pm$ ~0.001 (stat.) using signal MC. The resulting relative branching ratio is 
\[\frac{\brc(\lcpl \ra \Sigma^{0} \kpl)}{\brc(\lcpl \ra \Sigma^{0} \pipl )} = ~0.040~ \pm ~0.005 ~(\textnormal{stat.})~ \pm ~0.004 ~(\textnormal{syst.})~ .\]

\begin{figure}[!htb]
\begin{center}
\begin{tabular}{c}
\mbox{\includegraphics[height=7cm]{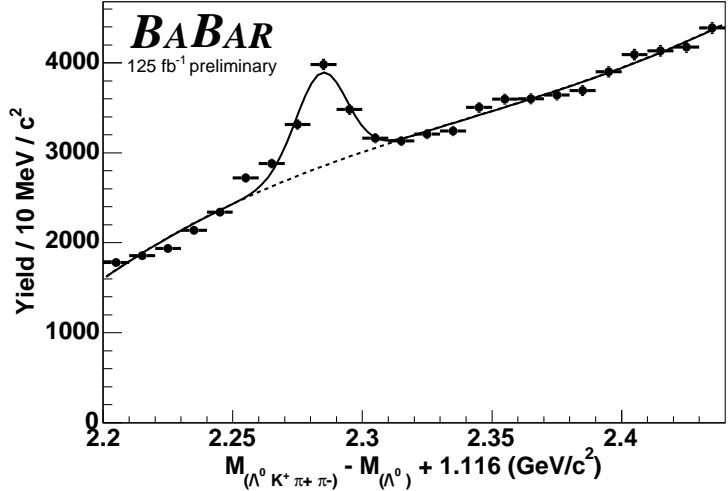}}
\end{tabular}
\caption{The invariant mass of  $\Lambda^{0}K^+\pi^{+}\pi^{-}$ combinations (GeV/$c^{2}$).}
\label{fig:lctolkpp}
\end{center}
\end{figure}
\subsection{Measurement of the decay $\Lambda^+_c \to \Lambda^{0}K^+\pi^{+}\pi^{-}$ }
~~~To measure the Cabibbo-suppressed decay $\Lambda^+_c \to \Lambda^{0}K^+\pi^{+}\pi^{-}$ we used the selection criteria described in Sec.~\ref{sec:Analysis}, but with scaled momentum restricted to $x_{p} >$ 0.6. The invariant mass distribution of $\Lambda^{0}K^+\pi^{+}\pi^{-}$ combinations is shown in Fig.~\ref{fig:lctolkpp}. The
 distribution is fitted with a Gaussian function for the signal shape and a third-order polynomial for background. We obtained the width $\sigma = 9.9 \pm 1.0~\mevcsq$ and signal yield of 2591 $\pm$ 258 (stat.) events for the $\Lambda^+_c \to \Lambda^{0}K^+\pi^{+}\pi^{-}$ decay with 10.1$\sigma$ statistical significance.
 A small fluctuation is seen at the mass 2.26 GeV/$c^2$. The effect of this fluctuation on the signal yield has been included in the fitting systematic. For normalization
 mode we use $\Lambda^+_c \to \Lambda^{0}\pi^{+}$ with scaled momentum at $x_{p} >$ 0.6, for which we obtained a raw yield of 22173 $\pm$ 287 (stat.) events. The relative signal 
reconstruction efficiency is measured to be $\epsilon ( \Lambda^+_c \to \Lambda^{0}K^{+}\pi^{+}\pi^{-})$ / $\epsilon ( \Lambda^+_c \to \Lambda^{0}\pi^{+})$ = 0.442 ~$\pm$~ 0.004 (stat.). The resulting branching ratio is 
\[\frac{\brc(\lcpl \ra \lz \kpl \pipl \pimi\ )}{\brc(\lcpl \ra \lz \pipl )} = ~0.266~ \pm ~0.027 ~(\textnormal{stat.})~ \pm ~0.032 ~(\textnormal{syst.}).~\]
This is the first measurement of the Cabibbo-suppressed decay $\Lambda^+_c \to \Lambda^{0}K^+\pi^{+}\pi^{-}$.
\newline

\subsection{Search for the decay of $\Lambda^+_c \to \Sigma^{0}K^+\pi^{+}\pi^{-}$}
\begin{figure}
\begin{center}
\begin{tabular}{c}
\mbox{\includegraphics[height=7cm]{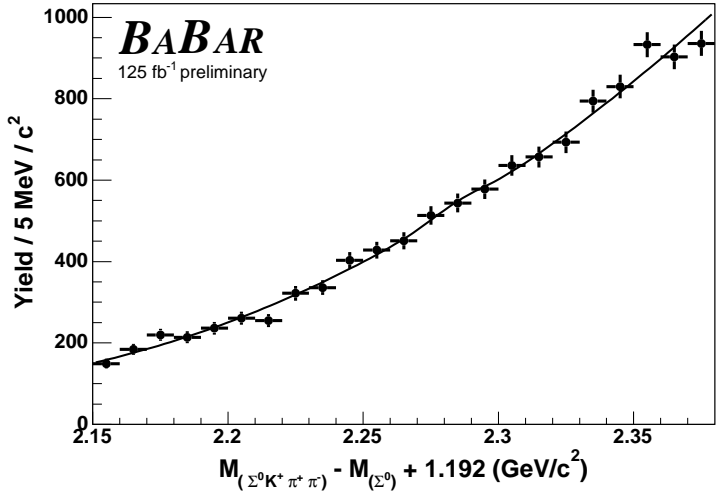}}
\end{tabular}
\caption{The invariant mass of $\Sigma^{0}K^+\pi^{+}\pi^{-}$ combinations (GeV/$c^{2}$).}
\label{fig:lctoskpp}
\end{center}
\end{figure}

~~~ We searched for the decay $\Lambda^+_c \to \Sigma^{0}K^+\pi^{+}\pi^{-}$ using the selection described in Sec.~\ref{sec:Analysis} and restricting the scaled momentum to $x_{p} >$ 0.6. The fit of the invariant mass distribution
 (Fig.~\ref{fig:lctoskpp}) yields 41 $\pm$ 51 (stat.) events 
for the $\Lambda^+_c \to \Sigma^{0}K^+\pi^{+}\pi^{-}$
 decay with 0.8$\sigma$ statistical significance.~The width $\sigma$ = 10.0~\mevcsq\ and mean = 2285.0~\mevcsq\ were fixed to values obtained from MC.
Using the decay mode $\Lambda^+_c \to \Sigma^{0}\pi^{+}$ for normalization, we find a raw yield of 8785 $\pm$ 131 (stat.) events for the decay $\Lambda^+_c \to \Sigma^{0}\pi^{+}$
 at $x_{p} >$ 0.6. The relative reconstruction efficiency is measured to be $\epsilon ( \Lambda^+_c \to \Sigma^{0}K^{+}\pi^{+}\pi^{-})$ / $\epsilon ( \Lambda^+_c \to \Sigma^{0}\pi^{+})$ = 0.390 ~$\pm$~0.002 (stat.). 
We do not observe any significant signal for $\Lambda^+_c \to \Sigma^{0}K^+\pi^{+}\pi^{-}$. Therefore, we calculate the upper limit using the Feldman and Cousins method~\cite{feld} including systematic uncertainties. We find:

\[\frac{\brc(\lcpl \ra \Sigma^{0} \kpl \pipl \pimi)}{\brc(\lcpl \ra \Sigma^{0} \pipl )} < 3.9 \times ~10^{-2}  ~@ ~90\% ~{\rm {CL}}.\]

\section{SYSTEMATIC STUDIES}
\label{sec:Systematics}
~~~We have considered several possible sources of systematic uncertainties in
 our measurements.
~The systematic uncertainty due to each cut used in 
candidate selection is $\sim$ 1 $\%$. The photon spectrum is 
different in measured and reference decay modes, so we studied the effect by changing 
the photon energy and assigned a systematic uncertainty of 2.8 $\%$ to the branching ratio ($ \Lambda^+_c \to \Sigma^{0}K^{+}$ / $ \Lambda^+_c \to \Sigma^{0}\pi^{+}$). 
\newline
\indent We have studied possible biases due to our fitting procedure. The shape of
background has been varied by changing the degree of the polynomial function as well
as the shapes of $\Sigma$ and $\Xi_c$ reflections (in case of $ \Lambda^+_c \to \Lambda^{0} \pi^{+}$ decay). Any change in signal yields is being
taken as a systematic uncertainty. We also varied the signal width ($\sigma$) by 
one standard deviation, with an change in yields interpreted as a systematic uncertainty.
We assign a systematic uncertainty due to fit bias of 5.1~$\%$ for $ \Lambda^+_c \to \Lambda^{0}K^{+}$ / $ \Lambda^+_c \to \Lambda^{0}\pi^{+}$, 8.0$~\%$ for $ \Lambda^+_c \to \Sigma^{0}K^{+}$ / $ \Lambda^+_c \to \Sigma^{0}\pi^{+}$, and 10.1$~\%$ for $ \Lambda^+_c \to \Lambda^{0}K^{+}\pi^{+}\pi^{-}$ / $  \Lambda^+_c \to \Lambda^{0}\pi^{+}$. The systematic uncertainty associated with the fitting is the dominant one.

\section{SUMMARY}
\label{sec:Summary}
We report on a measurement of the branching ratio of the Cabibbo-suppressed decays  $ \Lambda^+_c \to \Lambda^{0}K^{+}$ and $\Lambda^+_c \to \Sigma^{0}K^{+}$ with improved accuracy.
We also report the first 
observation of the Cabibbo-suppressed decay $\Lambda^+_c \to \Lambda^{0}K^+\pi^{+}\pi^{-}$, and we set an upper limit on the $\Lambda^+_c \to \Sigma^{0}K^+\pi^{+}\pi^{-}$ decay. The results for these decay modes are summarized in Table~\ref{tab:Summary}. All results reported in this paper are preliminary. 
The expectations from the quark model~\cite{Khana} are ${\brc(\lcpl \ra \lz \kpl)}/{\brc(\lcpl \ra \lz \pipl )} = [0.039-0.056]$ and ${\brc(\lcpl \ra \Sigma^{0} \kpl)}/{\brc(\lcpl \ra \Sigma^{0} \pipl )} = [0.033-0.036]$.
The results are in agreement with the predictions.

\begin{sidewaystable}
\begin{center}

\begin{tabular}{|l|c|c|c|c|c|c|}\hline\hline

  Signal Mode  & Signal  & Reference & Reference  & Relative    & 
\multirow{2}*{$\frac{\brc_{\rm signal}}{\brc_{\rm reference}}$} & Other Measurements \\  
                    & Yield   & Mode           & Yield       & Efficiency  &       &    \\ \hline

\rule[0.22in]{-0.05in}{0.0in} \lz \kpl\      & $1164 \pm 107$ &  \lz \pipl\ & $33594 \pm 367$ & 0.781 $\pm$ 0.004 & $0.044~ \pm ~0.004~ \pm ~0.002$ &   $0.074 \pm 0.0
10~ \pm ~ 0.012$~\cite{Belle}           \\ 
\rule[0.18in]{-0.05in}{0.0in}  \rule[-0.08in]{-0.05in}{0.0in}     &  &  &  &  & ($x_{p} > 0.5$) &($x_{p} > 0.5)$ \\ 
\hline

\rule[0.22in]{-0.05in}{0.0in}  \lz \kpl \pipl \pimi    & $2591 \pm 258$ & \lz \pipl\ & $22173 \pm 287$  & 0.442 $\pm$ 0.004 & $0.266~\pm~0.027~\pm~0.032$        &
   ---           \\ 
\rule[0.18in]{-0.05in}{0.0in}  \rule[-0.08in]{-0.05in}{0.0in}                         &                  &             &   &  & ($x_{p} > 0.6$) & \\ 
\hline

\rule[0.22in]{-0.05in}{0.0in} \sigz \kpl\ & $387 \pm 48$ & \sigz \pipl\ & $12450 \pm 170$ & 0.780 $\pm$ 0.001 & $0.040~ \pm ~0.005~ \pm ~0.004$ &   $0.056 \pm
 0.014~ \pm ~ 0.008$~\cite{Belle}           \\ 
\rule[0.18in]{-0.05in}{0.0in}  \rule[-0.08in]{-0.05in}{0.0in}                                &                   &                 &  & &($x_{p} > 0.5$)   & ($x_{p} > 0.6$) \\ 
\hline

\rule[0.22in]{-0.05in}{0.0in} \sigz \kpl \pipl \pimi  & $41 \pm 51$ & \sigz \pipl\ & $8785 \pm 131$  & 0.390 $\pm$ 0.002 & $< 3.9 \times ~10^{-2} ~ @ ~90
\%$ CL        &   ---           \\ 
\rule[0.18in]{-0.05in}{0.0in}  \rule[-0.08in]{-0.05in}{0.0in}                         &                  &             &  &  & ($x_{p} > 0.6$) & \\

\hline\hline
\end{tabular}
\end{center}
\caption{$\!\!$Summary of results obtained$\!$ in this analysis. $\!\!$The last column shows the current other measurements 
 of each decay mode, $\!$where $\!$ available.}
\label{tab:Summary}
\end{sidewaystable} 

\section{ACKNOWLEDGMENTS}
\label{sec:Acknowledgments}

\input pubboard/acknowledgements

\end{document}

%% file: includes/particle.tex
\def\babar{\mbox{\sl B\kern-0.1em{\smaller A}\kern-0.1em B\kern-0.1em{\smaller A\kern-0.2em R}}}

\RequirePackage{xspace}

\newcommand{\slac}{SLAC}

\newcommand{\brc}{\mbox{$\cal B$}}

\newcommand{\chitof}[1]{\mbox{$\chi_{c0}$}}

\newcommand{\mevcsq}{\mbox{\rm MeV$\!$/$c^2$}}
\def\invfb {\ensuremath{\mbox{\,fb}^{-1}}\xspace}

\newcommand{\pimi}{\mbox{$\pi^-$}}
\newcommand{\pipl}{\mbox{$\pi^+$}}

\newcommand{\p}{\mbox{$p$}}

\newcommand{\kpl}{\mbox{$K^+$}}

\newcommand{\phiz}{\mbox{$\phi$}}

\newcommand{\lz}{\mbox{$\Lambda^{0}$}}

\newcommand{\sigz}{\mbox{$\Sigma^0$}}

\newcommand{\lcpl}{\mbox{$\Lambda^+_c$}}

\newcommand{\dz}{\mbox{$D^0$}}

     \def\plb#1#2#3#4{#1, Phys.\ Lett.\ B {\bf#2}, #3 (#4)}
     \def\pr#1#2#3#4{#1, Phys.\ Rev.\ {\bf#2}, #3 (#4)}
     \def\prd#1#2#3#4{#1, Phys.\ Rev.\ D {\bf#2}, #3 (#4)}
     \def\prl#1#2#3#4{#1, Phys.\ Rev.\ Lett.\ {\bf#2}, #3 (#4)}
     
     \def\nim#1#2#3#4{#1, Nucl.\ Instrum.\ Methods\ {\bf#2}, #3 (#4)}
\def\zpc#1#2#3#4{#1, Z. Phys. C\ {\bf#2}, #3 (#4)}
     
\def\npb#1#2#3#4{#1, Nucl.\ Phys.\ {\bf B#2}, #3 (#4)}
     
\newcommand{\bcen}{\begin{center}}
\newcommand{\ecen}{\end{center}}
\newcommand{\bfl}{\begin{flushleft}}
\newcommand{\efl}{\end{flushleft}}
\newcommand{\bfr}{\begin{flushright}}
\newcommand{\efr}{\end{flushright}}
\newcommand{\bequ}{\begin{equation}}
\newcommand{\eequ}{\end{equation}}
\newcommand{\beqa}{\begin{eqnarray*}}
\newcommand{\eeqa}{\end{eqnarray*}}
\newcommand{\beq}{\begin{eqnarray}}
\newcommand{\eeq}{\end{eqnarray}}
\newcommand{\bitm}{\begin{itemize}}
\newcommand{\eitm}{\end{itemize}}



%% file: pubboard/authors_sum2004.tex
\begin{center}
\small

The \babar\ Collaboration,
\bigskip

%
B.~Aubert,
R.~Barate,
D.~Boutigny,
F.~Couderc,
J.-M.~Gaillard,
A.~Hicheur,
Y.~Karyotakis,
J.~P.~Lees,
V.~Tisserand,
A.~Zghiche
\inst{Laboratoire de Physique des Particules, F-74941 Annecy-le-Vieux, France }
% ...

\end{center}\newpage

%% file: pubboard/acknowledgements.tex
We are grateful for the 
extraordinary contributions of our \pep2\ colleagues in
achieving the excellent luminosity and machine conditions
that have made this work possible.
The success of this project also relies critically on the 
expertise and dedication of the computing organizations that 
support \babar.
The collaborating institutions wish to thank 
SLAC for its support and the kind hospitality extended to them. 
This work is supported by the
US Department of Energy
and National Science Foundation, the
Natural Sciences and Engineering Research Council (Canada),
Institute of High Energy Physics (China), the
Commissariat \`a l'Energie Atomique and
Institut National de Physique Nucl\'eaire et de Physique des Particules
(France), the
Bundesministerium f\"ur Bildung und Forschung and
Deutsche Forschungsgemeinschaft
(Germany), the
Istituto Nazionale di Fisica Nucleare (Italy),
the Foundation for Fundamental Research on Matter (The Netherlands),
the Research Council of Norway, the
Ministry of Science and Technology of the Russian Federation, and the
Particle Physics and Astronomy Research Council (United Kingdom). 
Individuals have received support from 
CONACyT (Mexico),
the A. P. Sloan Foundation, 
the Research Corporation,
and the Alexander von Humboldt Foundation.